\documentclass[%
 reprint,
 amsmath,amssymb,
 aps,
]{revtex4-1}

\usepackage{graphicx}
\usepackage{dcolumn}
\usepackage{bm}

\begin{document}

\preprint{APS/123-QED}

\title{Large Non-Gaussianity in Non-Minimally Coupled Derivative Inflation with Gauss-Bonnet Correction }

\author{Kourosh Nozari}
 \homepage{knozari@umz.ac.ir}
\author{Narges Rashidi}
\homepage{n.rashidi@umz.ac.ir}%
\affiliation{Department of Physics, Faculty of Basic Sciences,\\
University of Mazandaran,\\
P. O. Box 47416-95447, Babolsar, IRAN}

\date{\today}

\begin{abstract}
We study a nonminimal derivative inflationary model
in the presence of the Gauss-Bonnet term. To have a complete
treatment of the model, we consider a general form of the nonminimal
derivative function and also the Gauss-Bonnet coupling term. By
following the ADM formalism, expanding the action up to the third order
in the perturbations and using the correlation functions, we study
the perturbation and its non-Gaussian feature in details. We also
study the consistency relation that gets modified in the presence
of the Gauss-Bonnet term in the action. We compare the results of our
consideration in confrontation with Planck2015 observational data and
find some constraints on the model's parameters. Our treatment shows that this
model in some ranges of the parameters is consistent with the
observational data. Also, in some ranges of model's parameters, the
model predicts blue-tilted power spectrum. Finally, we show that nonminimal derivative
 model in the presence of the GB term has capability to have large non-Gaussianity.
\begin{description}
\item[PACS numbers]
04.50.Kd , 98.80.Cq , 98.80.Es
\item[Key Words]
Inflation, Cosmological Perturbations, Non-Gaussianity, Nonminimal Derivative
Coupling, Observational Constraints.
\end{description}
\end{abstract}

\maketitle


\section{Introduction}

Inflation is accepted as a paradigm of the early Universe to solve
some problems of the standard big bang cosmology, such as the
flatness, horizon and relics problems. An inflationary paradigm, to
be a successful paradigm, should also provide a causal graceful
mechanism for generating the primordial density perturbations needed
to seed the formation of structures in the Universe
~\cite{Gut81,Lin82,Alb82,Lin90,Lid00a,Lid97,Rio02,Lyt09}. A simple
inflationary scenario involves a single scalar field which its
nearly flat potential dominated the energy density of the early
Universe. Such a simple inflation model predicts the dominant mode of
the primordial density fluctuations to be almost adiabatic, scale
invariant and also Gaussian distributed~\cite{Mal03}. However, recently released
Planck observational data have detected a level of scale
dependence in the primordial density
perturbations~\cite{Ade15a,Ade15b}. Also, Planck collaboration has
obtained some constraints on the primordial non-Gaussianity in the
perturbations~\cite{Ade15c}. Some authors have proposed extended
inflationary models in light of the scalar-tensor theories which predict
a level of non-Gaussianity in the density perturbation's
mode~\cite{Mal03,Bar04,Che10,Fel11a,Fel11b,Noz12,Noz13a,Noz13b,Noz13c}.
Prediction of non-Gaussianity in the theory is really an important and impressive
point, in the sense that a large amount of information about the
cosmological dynamics deriving the initial inflationary expansion is
carried by the non-Gaussian perturbations~\cite{Bab04a,Che08}. In
this respect, it seems that such extended inflationary models are
more favorable.

One successful class of the extended inflationary models in light of
scalar-tensor theories, proposed by Amendola, is the one with a
nonminimal coupling between the inflaton kinetic term and Einstein
tensor~\cite{Ame93}. By adopting such a coupling, the friction is
enhanced gravitationally at higher energies, leading to increase in the
friction of an inflaton rolling down its own potential. Coupling
between the inflaton kinetic term and Einstein tensor which is
usually given by the lagrangian term as
$\left(R_{\mu\nu}-\frac{1}{2}g_{\mu\nu}R\right)\nabla^{\mu} \phi
\nabla^{\nu}\phi$, satisfies the unitary invariance of the theory
during inflation~\cite{Ger10}. Authors have studied the early time
accelerating expansion of the universe as well as the late time
dynamics by considering the model with coupling between the
derivatives of the scalar field and gravitational
sector~\cite{Tsu12,Sad2013,Sar10,Noz15,Noz14}.

Thinking over the very early Universe approaching the Planck scale,
we could consider some quantum corrections on Einstein gravity. This
means that Einstein gravity is a low-energy limit of a quantum
theory of gravity. The most promising candidate for quantum gravity
is string theory. String theory suggests that one way to accommodate
quantum effects of gravity is to include quadratic curvature terms
in a gravitational action as $R_{abcd}R^{abcd}-4R_{ab}R^{ab}+R^{2}$.
The presence of a Gauss-Bonnet combination in the action does not
have any ghost as well as any problem with the unitarity. Such a
quadratic term plays a significant role in the early Universe
dynamics~\cite{Zwi85,Bou85}. However it turns out that the
Gauss-Bonnet term in dimensions less than five, is a topological
term having no influence in the background dynamics. In this regard,
to consider the effect of the Gauss-Bonnet term on the field
evolution even in four dimension, we should introduce a coupling
between a scalar field and the Gauss-Bonnet term as the effective
theory added to quantum correction
\cite{Noj05,Noj07,Guo09,Guo10}(for higher dimensional extensions
we refer to
~\cite{Bro07,Bam07,And07,Noz08,Noz09a,Noz09b,Noz09c}).

In this paper, we study cosmological dynamics of a generalized
inflationary model in which both the scalar field and its derivative
are coupled to the curvature. The scalar field is coupled to the
Gauss-Bonnet term and its derivative is coupled to the Einstein
tensor. These couplings have no problem with unitarity. In this
work, we consider a general form of the nonminimal derivative term
in the lagrangian which includes the simple cases studied in
literatures. The extended form of the nonminimal derivative term is
given by ${\cal{N}}(\phi)G_{\mu\nu}\nabla^{\mu}\nabla^{\nu}(\phi)$,
in which, ${\cal{N}}$ is a general function of $\phi$. This
expression, for ${\cal{N}}\sim\frac{1}{2}\phi$, leads to the simple
nonminimal derivative case.

To explore the viability of an inflationary model, it is important
to study properties of the initial cosmological perturbations.
Some parameters, such as the scalar spectral index and tensor-to-scalar
ratio are important and helpful parameters to describe the main
properties of the cosmological perturbations. By comparing the calculated values
of the mentioned parameters in the model at hand with the observational
values, we can explore whether the inflationary model is
successful and viable or not. In this respect, several collaborations have
tried to obtain some observational constraints on these parameters.
From the joint WMAP9+eCMB+BAO+H$_{0}$ data, WMAP collaboration
has obtained $r<0.13$ and $n_{s}=0.9636\pm 0.0084$~\cite{Hin13}.
Planck collaboration, by using the joint Planck 2013+WMAP9+BAO data
has constrained the scalar spectral index and tensor-to-scalar ratio
as $r<0.12$ and $n_{s}=0.9643\pm 0.0059$~\cite{Ade13}. The new study
of Planck team has released the constraints on the perturbative
parameters as $r<0.099$ and $n_{s}= 0.9652\pm 0.0047$, from Planck TT, TE,
EE+lowP +WP data. It should be noted that Planck TT, TE, EE+lowP
refer to the combination of the likelihood at $l>30$ using TT, TE,
and EE spectra and the low $l$ multipole temperature polarization
likelihood~\cite{Ade15a,Ade15b,Ade15c}.

With these points in mind, in section II we present the main equations of our
extended inflationary model. In section III, by using the ADM
formalism, we study the linear perturbations of the model at hand.
In this regard, we expand the action of the model up to the second
order in perturbation and consider the 2-point correlator to explore
the amplitude of the scalar perturbation and also its spectral
index. As well as, the tensor perturbation and its spectral index
are obtained by exploring the tensor part of the perturbed metric.
By considering the third order action, the non-linear perturbation
in our setup is studied in section IV. In this section, by using the
3-point correlator, the non-Gaussian feature of the primordial
perturbation is obtained. In the limit $k_{1}=k_{2}=k_{3}$, we
compute the equilateral and orthogonal configuration of the
non-Gaussianity. In section V, we examine our generalized inflationary
model in confrontation with Planck2015 observational data and obtain
the domain of the parameters which makes the model observationally
viable.

\section{The Setup}
The four-dimensional action for a cosmological model in the presence
of a nonminimal coupling between the inflaton and the Gauss-bonnet
term and a nonminimal coupling between the derivatives of the
inflaton and Einstein tensor, is expressed as
\begin{eqnarray}
\label{eq1} S=\int
d^{4}x\sqrt{-g}\Bigg[\frac{1}{2\kappa^{2}}R-\frac{1}{2}\partial_{\mu}\phi
\partial^{\mu}\phi-V(\phi)-\alpha(\phi){\cal{L}_{GB}}\nonumber\\
+{\cal{N}}(\phi)G_{\mu\nu}\nabla^{\mu}\nabla^{\nu}(\phi)\Bigg]\,,\hspace{0.6cm}
\end{eqnarray}
where, $R$ is the Ricci scalar, $\phi$ is an inflaton filed followed
by the potential $V(\phi)$. $\alpha(\phi)$ and ${\cal{N}}(\phi)$ are
general functions of the scalar field which show the nonminimal
couplings. ${\cal{L}_{GB}}$ is the lagrangian of the Gauss-Bonnet
term. Also, $G_{\mu\nu}$ is Einstein tensor defined as
$G_{\mu\nu}=R_{\mu\nu}-\frac{1}{2}g_{\mu\nu}R$. In some papers, one
see a coefficient $\frac{\kappa^{*}}{2}$, with dimension of
length-squared, in front of the nonminimal derivative term. In this
work, we absorb such a coefficient in nonminimal derivative function
${\cal{N}}(\phi)$. Varying the action \eqref{eq1} with respect to
the metric, by assuming the FRW metric, leads to the following
Friedmann equation of the model
\begin{equation}
\label{eq2}
H^{2}=\frac{\kappa^{2}}{3}\left(\dot{\phi}^{2}\Big(\frac{1}{2}-9H^{2}{\cal{N}}'\Big)+V(\phi)+24H^{3}\dot{\alpha}\right),
\end{equation}
where, a dot denotes a time derivative of the parameter and a prime
refers to a derivative with respect to the inflaton field. Variation
of the action \eqref{eq1} with respect to the scalar field, gives
the equation of motion of the inflaton as follows
\begin{eqnarray}
\label{eq3}
\ddot{\phi}\Big(-1+6{\cal{N}}'H^{2}\Big)+\Big(12{\cal{N}}'H\dot{H}+18{\cal{N}}'H^{3}-3H\Big)\dot{\phi}\nonumber\\
+6{\cal{N}}''H^{2}\dot{\phi}^{2}-24H^{4}\alpha'-24H^{2}\dot{H}\alpha'-V'=0\,.\hspace{0.6cm}
\end{eqnarray}

In this extended inflationary model, the slow-roll parameters
defined as $\epsilon\equiv-\frac{\dot{H}}{H^{2}}$ and
$\eta=-\frac{1}{H}\frac{\ddot{H}}{\dot{H}}$, are given by the
following expressions
\begin{equation}
\label{eq4} \epsilon=\frac{E}{1+\kappa^{2}
{\cal{N}}'\dot{\phi}^{2}-8\kappa^{2}H\dot{\alpha}},
\end{equation}
\begin{eqnarray}
\label{eq5} \eta=2\epsilon-\frac{\dot{E}}{H\epsilon(1+\kappa^{2}
{\cal{N}}'\dot{\phi}^{2}-8\kappa^{2}H\dot{\alpha})}\nonumber\\
+\frac{E}{H\epsilon}\frac{\kappa^{2}{\cal{N}}''\dot{\phi}^{3}+2\kappa^{2}{\cal{N}}'\ddot{\phi}\dot{\phi}}
{(1+\kappa^{2}
{\cal{N}}'\dot{\phi}^{2}-8\kappa^{2}H\dot{\alpha})^{2}}
\nonumber\\
+\frac{E}{H\epsilon}\frac{-8\kappa^{2}H\alpha''
\dot{\phi}^{2}-8\kappa^{2}H\alpha'\ddot{\phi}-8\dot{H}\alpha'\dot{\phi}}
{(1+\kappa^{2}
{\cal{N}}'\dot{\phi}^{2}-8\kappa^{2}H\dot{\alpha})^{2}},
\end{eqnarray}
where parameter $E$ is defined as
\begin{eqnarray}
\label{eq6} E\equiv
\frac{2\kappa^{2}{\cal{N}}'\dot{\phi}\ddot{\phi}}{H}+
\frac{\kappa^{2}\dot{\phi}^{2}}{2H^{2}}-3\kappa^{2}{\cal{N}}'\dot{\phi}^{2}
+\frac{\kappa^{2}{\cal{N}}''\dot{\phi}^{3}}{2H}-4\kappa^{2}\ddot{\alpha}\nonumber\\
+4\kappa^{2}H\dot{\phi}\alpha'\,.\hspace{0.5cm}
\end{eqnarray}
The evolution of the Hubble parameter during the inflationary era is
so slow, satisfying the conditions $\epsilon \ll 1$ and $\eta \ll
1$. As one of these two slow-varying parameters reaches unity, the
inflation phase of the early universe terminates.

The slow-roll limits in the simple single field inflationary model
are given by $\ddot{\phi}\ll |3H\dot{\phi}|$ and $\dot{\phi}^{2}\ll
V(\phi)$. In a model with NMDC and GB terms these conditions get modified.
By considering the high friction regime
(${\cal{N}}'H^{2}\geq1$), the slow-roll conditions in this setup are
$8\kappa^{2}H|\dot{\alpha}|\ll 1$, $|\ddot{\alpha}|\ll
|\dot{\alpha}H|$, $9H^{2}|{\cal{N}}'|\dot{\phi}^{2}\ll V$ and
$|\ddot{{\cal{N}}}|\ll |3\dot{{\cal{N}}}H^{3}|$ (the slow roll conditions in the
presence of the GB term are discussed in \cite{Guo10,Car15}). By
using these conditions and considering $\dot{H}\ll H^{2}$ we can
rewrite the Friedmann equation and equation of motion as follows
\begin{equation}
\label{eq2-2} H^{2}\simeq \frac{\kappa^{2}}{3}V
\end{equation}
\begin{equation}
\label{eq3-2} \Big(18{\cal{N}}'H^{3}-3H\Big)\dot{\phi}
-24H^{4}\alpha'-V'\simeq0\,.\hspace{0.6cm}
\end{equation}
The number of e-folds during inflationary era which is defined as
\begin{equation}
\label{eq7} N=\int_{t_{hc}}^{t_{f}} H dt\,,
\end{equation}
in our extended setup and within the slow-roll limits is given by
the following expression
\begin{equation}
\label{eq8} N\simeq \int_{\phi_{hc}}^{\phi_{f}} \frac{3H^{2}
\Big(6{\cal{N}}'H^{2}-1\Big)}{V'+24H^{4}\alpha'} d\phi\,,
\end{equation}
In equation \eqref{eq8}, we have shown the value of the inflaton
filed at the time of the horizon crossing of the universe scale by
$\phi_{hc}$ and the value of the field at the time of exit from
inflationary phase by $\phi_{f}$.

We proceed by studying the linear perturbation of the model
in the next section. To explore the linear perturbations, we
calculate the spectrum of perturbations which are produced due to
quantum fluctuations of the fields about their homogeneous
background values.

\section{Linear Perturbations}
Quantum behavior of both the scalar field $\phi$ and the
space-time metric, $g_{\mu\nu}$, around their homogeneous background
solution, contribute in the perturbation. In this section, we study
the linear perturbation arising from fluctuation of both the scalar
field and the space-time metric. To study the linear perturbation,
the first step is expanding the action \eqref{eq1} up to the second
order in small fluctuations. To this end, it is convenient to work
in ADM metric formalism given by~\cite{Arn60}
\begin{equation}
\label{eq9}
ds^{2}=-N^{2}dt^{2}+h_{ij}\big(dx^{i}+N^{i}dt\big)\big(dx^{j}+N^{j}dt\big),
\end{equation}
In this metric,  $N^{i}$ is the shift vector and $N$ is the lapse
function. By expanding the shift and laps functions as $N^{i}\equiv
\Phi^{i}=\delta^{ij}\partial_{j}\Phi+v^{i}$ and $N=1+\Psi$, we
obtain a general perturbed form of the metric \eqref{eq9}. In this
expansion, $\Psi$ and $\Phi$ are 3-scalars and $v^{i}$ is a vector
which satisfies the condition $v^{i}_{,i}=0$~\cite{Muk92,Bau09}. The
coefficient $h_{ij}$ is written as
$h_{ij}=a^{2}\left[(1-2{\Upsilon})\delta_{ij}+2{\cal{T}}_{ij}\right]$,
with ${\Upsilon}$ as the spatial curvature perturbation and
${\cal{T}}_{ij}$ as a spatial shear 3-tensor. Note that
${\cal{T}}_{ij}$ is a symmetric and traceless tensor. Now, the
perturbed metric \eqref{eq9} is written as the following form
\begin{eqnarray}
\label{eq10} ds^{2}=
-(1+2\Psi)dt^{2}+2a(t)\Phi_{i}\,dt\,dx^{i}\hspace{1.5cm}\nonumber\\
+a^{2}(t)\left[(1-2{\Upsilon})\delta_{ij}+2{\cal{T}}_{ij}\right]dx^{i}dx^{j}.
\end{eqnarray}
Since we are going to study the scalar perturbation of the theory,
it is convenient to work within the uniform-field gauge in which
$\delta\phi=0$ and also the gauge ${\cal{T}}_{ij}=0$. Now,
by considering the scalar part of the perturbations at the linear level
and within the uniform-field gauge, the perturbed metric is given as
follows ~\cite{Muk92,Bau09,Bar80}
\begin{eqnarray}
\label{eq11}
ds^{2}=-(1+2\Psi)dt^{2}+2a(t)\Phi_{,i}\,dt\,dx^{i}\hspace{1.5cm}\nonumber\\
+a^{2}(t)(1-2{\Upsilon})\delta_{ij}dx^{i}dx^{j}.
\end{eqnarray}
We replace the perturbed metric \eqref{eq11} in action \eqref{eq1},
expand the action up to the second order in perturbations and obtain
the following expression
\begin{eqnarray}
\label{eq12} S_{2}=\int dt\,d^{3}x\,
a^{3}\Bigg[-3(\kappa^{-2}+\dot{\phi}^{2}{\cal{N}}'-8H\dot{\alpha})\dot{\Upsilon}^{2}-\hspace{1cm}\nonumber\\
\frac{2(\kappa^{-2}+\dot{\phi}^{2}{\cal{N}}'-8H\dot{\alpha})}{a^{2}}\Psi
\partial^{2}{\Upsilon}+\frac{1}{a^{2}}\Big(2(\kappa^{-2}+\dot{\phi}^{2}{\cal{N}}'
\nonumber\\
-8H\dot{\alpha})\dot{\Upsilon}-\big(2\kappa^{-2}H+6H\dot{\phi}^{2}{\cal{N}}'-24H^{2}\dot{\alpha}\big)\Psi\Big)\partial^{2}\Phi\nonumber\\
+3\Big(2\kappa^{-2}H+6H\dot{\phi}^{2}{\cal{N}}'-24H^{2}\dot{\alpha}\Big)\Psi\dot{\Upsilon}
\nonumber\\
-\Big(3\kappa^{-2}H^{2}-\frac{1}{2}\dot{\phi}^{2}+18H^{2}\dot{\phi}^{2}{\cal{N}}'-48H^{3}\dot{\alpha}\Big)\Psi^{2}\nonumber\\
+\frac{\kappa^{-2}+\dot{\phi}^{2}{\cal{N}}'-8H\dot{\alpha}}{a^{2}}(\partial{\Upsilon})^{2}\Bigg]\,.\hspace{1cm}
\end{eqnarray}
By varying the action \eqref{eq12} in this gauge, we get the
following expressions for the equations of motion of $\Psi$ and
$\Phi$
\begin{equation}
\label{eq13}
\Psi=2\frac{\kappa^{-2}+\dot{\phi}^{2}{\cal{N}}'-8H\dot{\alpha}}{2\kappa^{-2}H+6H\dot{\phi}^{2}{\cal{N}}'-24H^{2}\dot{\alpha}}\dot{\Upsilon},
\end{equation}
\begin{eqnarray}
\label{eq14}
\frac{\partial^{2}\Phi}{a^{2}}=3\dot{\Upsilon}-\frac{\kappa^{-2}+\dot{\phi}^{2}{\cal{N}}'-8H\dot{\alpha}}{a^{2}(2\kappa^{-2}H+
6H\dot{\phi}^{2}{\cal{N}}'-24H^{2}\dot{\alpha})}
\partial^{2}{\Upsilon}\nonumber\\-\frac{2(3\kappa^{-2}H^{2}-\frac{1}{2}\dot{\phi}^{2}
+18H^{2}\dot{\phi}^{2}{\cal{N}}'-48H^{3}\dot{\alpha})}{(2\kappa^{-2}H+6H\dot{\phi}^{2}{\cal{N}}'-24H^{2}\dot{\alpha})}\Psi. \nonumber\\
\end{eqnarray}
We substitute equation of motion \eqref{eq13} in the second order
action \eqref{eq12} and integrate it by parts to find
\begin{equation}
\label{eq15} S_{2}=\int
dt\,d^{3}x\,a^{3}{\cal{W}}\left[\dot{\Upsilon}^{2}-\frac{c_{s}^{2}}{a^{2}}(\partial
{\Upsilon})^{2}\right],
\end{equation}
where the parameters ${\cal{W}}$ and $c_{s}^{2}$ (known as the sound
velocity) are defined as
\begin{widetext}
\begin{eqnarray}
\label{eq16}
{\cal{W}}=-4\frac{\left(\kappa^{-2}+\dot{\phi}^{2}{\cal{N}}'-8H\dot{\alpha}\right)^{2}\left(9\kappa^{-2}H^{2}-\frac{3}{2}\dot{\phi}^{2}
+54H^{2}\dot{\phi}^{2}{\cal{N}}'-144H^{3}\dot{\alpha}\right)}{3\left(
2\kappa^{-2}H+6H\dot{\phi}^{2}{\cal{N}}'-24H^{2}\dot{\alpha}\right)^{2}}
+3\left(\kappa^{-2}+\dot{\phi}^{2}{\cal{N}}'-8H\dot{\alpha}\right),
\end{eqnarray}
\end{widetext}

\begin{widetext}
\begin{eqnarray}
\label{eq17}
c_{s}^{2}=3\Bigg[2\Big(2\kappa^{-2}H+6H\dot{\phi}^{2}{\cal{N}}'-24H^{2}\dot{\alpha}\Big)\Big
(\kappa^{-2}+\dot{\phi}^{2}{\cal{N}}'-8H\dot{\alpha}\Big)H
-\Big(2\kappa^{-2}H+6H\dot{\phi}^{2}{\cal{N}}'-24H^{2}\dot{\alpha}\Big)^{2}
\nonumber\\
\Big(\kappa^{-2}+\dot{\phi}^{2}{\cal{N}}'-8H\dot{\alpha}\Big)^{-1}\Big(\kappa^{-2}-\dot{\phi}^{2}{\cal{N}}'-8\ddot{\alpha}\Big)\nonumber\\
+4\Big(2\kappa^{-2}H+6H\dot{\phi}^{2}{\cal{N}}'-24H^{2}\dot{\alpha}\Big)
\frac{d\Big(\kappa^{-2}+\dot{\phi}^{2}{\cal{N}}'-8H\dot{\alpha}\Big)}{dt}
\nonumber\\
-2\Big(\kappa^{-2}+\dot{\phi}^{2}{\cal{N}}'-8H\dot{\alpha}\Big)\,
\frac{d(2\kappa^{-2}H+6H\dot{\phi}^{2}{\cal{N}}'-24H^{2}\dot{\alpha})}{dt}\Bigg]
\Bigg[ \Bigg(9\Big(2\kappa^{-2}H+6H\dot{\phi}^{2}{\cal{N}}'-24H^{2}\dot{\alpha}\Big)^{2}\nonumber\\
-4\Big(\kappa^{-2}+\dot{\phi}^{2}{\cal{N}}'-8H\dot{\alpha}\Big)\Big(9\kappa^{-2}H^{2}-\frac{3}{2}\dot{\phi}^{2}
+54H^{2}\dot{\phi}^{2}{\cal{N}}'-144H^{3}\dot{\alpha}\Big)
\Bigg)\Bigg]^{-1},
\end{eqnarray}
\end{widetext}
See Refs.~\cite{Fel11a,Fel11b,Che08,See05} for more details in
driving this type of equations.

To proceed, we calculate the quantum perturbations of ${\Upsilon}$.
In this regard, we obtain the equation of motion of ${\Upsilon}$ by
varying the action \eqref{eq15}. The result becomes
\begin{equation}
\label{eq18}
\ddot{\Upsilon}+\left(3H+\frac{\dot{\cal{W}}}{\cal{W}}\right)\dot{\Upsilon}+c_{s}^{2}\,\frac{k^{2}}{a^{2}}\,{\Upsilon}=0.
\end{equation}
Up to the lowest order of the slow-roll approximation, solving the
above equation gives
\begin{equation}
\label{eq19}
{\Upsilon}=\frac{iHe^{-ic_{s}k\tau}}{2c_{s}^{\frac{3}{2}}\sqrt{k^{3}{\cal{W}}}}\left(1+ic_{s}k\tau\right).
\end{equation}
To survey the power spectrum of the curvature perturbation, it is
needed to obtain the two point correlation function in our setup.
Finding the vacuum expectation value of ${\Upsilon}$ at $\tau=0$,
which is corresponding to the end of inflation phase, gives the
two-point correlation function as
\begin{equation}
\label{eq20} \langle
0|{\Upsilon}(0,\textbf{k}_{1}){\Upsilon}(0,\textbf{k}_{2})|0\rangle
=(2\pi)^{3}\delta^{3}(\textbf{k}_{1}+\textbf{k}_{2})\frac{2\pi^{2}}{k^{3}}{\cal{A}}_{s},
\end{equation}
where ${\cal{A}}_{s}$, dubbed the power spectrum, is given by
\begin{equation}
\label{eq21}
{\cal{A}}_{s}=\frac{H^{2}}{8\pi^{2}{\cal{W}}c_{s}^{3}}\,.
\end{equation}
The scalar spectral index of the perturbations, which gives the
scale dependence of the perturbation, at the time of the Hubble
crossing is defined as
\begin{equation}
\label{eq22} n_{s}-1=\frac{d \ln {\cal{A}}_{s}}{d \ln
k}\Bigg|_{c_{s}k=aH},
\end{equation}
To find the scalar spectral index in our setup, we should notice
that equations \eqref{eq16} and \eqref{eq17} give
\begin{equation}
\label{eq22-2}
\kappa^{2}{\cal{W}}c_{s}^{2}=\epsilon-4\kappa^{2}H\dot{\alpha}+{\cal{O}}(\epsilon^{2})\,.
\end{equation}
So, from equation \eqref{eq21} in the linear order we find
\begin{equation}
\label{eq22-3}
{\cal{A}}_{s}=\frac{H^{2}}{8\pi^{2}(\epsilon-4\kappa^{2}H\dot{\alpha})c_{s}}\,.
\end{equation}
Since $d\ln k\equiv dN=H dt$, by using equations \eqref{eq22} and
\eqref{eq22-3} we get
\begin{eqnarray}
\label{eq23} n_{s}-1= -2\epsilon-\frac{1}{H}\frac{d \ln
(\epsilon-4\kappa^{2}H\alpha'\dot{\phi})}{dt}\nonumber\\
-\frac{1}{H}\frac{d \ln c_{s}}{dt}\,.\hspace{1cm}
\end{eqnarray}
The scale dependence of the perturbation is shown by deviation of
$n_{s}$ from the unity.

Some information about the initial perturbation can be found by
exploring the amplitude of the tensor perturbation and its spectral
index. We should use the tensor part of the perturbed metric
\eqref{eq10} to study the tensor perturbations. In terms of the two
polarization tensors, we write ${\cal{T}}_{ij}$ as follows
\begin{equation}
\label{eq24}
{\cal{T}}_{ij}={\cal{T}}_{+}\vartheta_{ij}^{+}+{\cal{T}}_{\times}\vartheta_{ij}^{\times},
\end{equation}
In equation \eqref{eq24} $\vartheta_{ij}^{(+,\times)}$ are
symmetric, traceless and transverse tensors. From the normalization
condition we have the following constraints
\begin{equation}
\label{eq25}
\vartheta_{ij}^{(+,\times)}(\textbf{k})\,\vartheta_{ij}^{(+,\times)}(-\textbf{k})^{*}=2,
\end{equation}
\begin{equation}
\label{eq26}
\vartheta_{ij}^{(+)}(\textbf{k})\,\vartheta_{ij}^{(\times)}(-\textbf{k})^{*}=0,
\end{equation}
and from the reality condition we have
\begin{equation}
\label{eq27}
\vartheta_{ij}^{(+,\times)}(-\textbf{k})=\left(\vartheta_{ij}^{(+,\times)}(\textbf{k})\right)^{*}.
\end{equation}
Now, the second order action for the tensor mode or gravitational
waves becomes as follows
\begin{eqnarray}
\label{eq28} S_{T}=\int dt\, d^{3}x\, a^{3}
{\cal{W}}_{T}\left[\dot{\cal{T}}_{+}^{2}-\frac{c_{\cal{T}}^{2}}{a^{2}}(\partial
{\cal{T}}_{+})^{2}+\dot{\cal{T}}_{\times}^{2}-\frac{c_{\cal{T}}^{2}}{a^{2}}(\partial
{\cal{T}}_{\times})^{2}\right],\nonumber\\
\end{eqnarray}
with the parameters ${\cal{W}}_{T}$ and $c_{T}^{2}$ defined as
\begin{equation}
\label{eq29}
{\cal{W}}_{T}=\frac{1}{4\kappa^{2}}\left(1+\kappa^{2}{\cal{N}}'\dot{\phi}^{2}-8\kappa^{2}H\alpha'\dot{\phi}\right),
\end{equation}
\begin{equation}
\label{eq30}
c_{T}^{2}=1-2\kappa^{2}{\cal{N}}'\dot{\phi}^{2}+8\kappa^{2}H\alpha'\dot{\phi}\,.
\end{equation}
To find the amplitude of the tensor perturbations we follow the
strategy applied for the scalar mode case and obtain
\begin{equation}
\label{eq31}
{\cal{A}}_{T}=\frac{H^{2}}{2\pi^{2}{\cal{W}}_{\cal{T}}c_{T}^{3}}.
\end{equation}
By using equation \eqref{eq31} and the definition of the tensor
spectral index
\begin{equation}
\label{eq32} n_{T}=\frac{d \ln {\cal{A}}_{T}}{d \ln k},
\end{equation}
we get the following expression for $n_{T}$ up to leading order
\begin{equation}
\label{eq33} n_{T}=-2\epsilon\,.
\end{equation}
Another important perturbation parameter which gives some
information on the perturbation, is the ratio between the amplitudes
of the tensor and scalar perturbations (briefly, tensor-to-scalar
ratio). This parameter in our setup is given by the following
expression
\begin{equation}
\label{eq34}
r=\frac{{\cal{A}}_{T}}{{\cal{A}}_{s}}\simeq16c_{s}\left(\epsilon-4\kappa^{2}H\alpha'\dot{\phi}\right).
\end{equation}

This equation shows that the presence of the extended NMDC term,
preserves the standard form of the consistency relation as
$r=-8c_{s}n_{T}$. Whereas, the presence of the GB term modifies the
consistency relation as equation \eqref{eq34}.

\section{Nonlinear Perturbations and Non-Gaussianity}

Another important aspect of an inflationary model is given by
non-Gaussianity of the primordial density perturbations. For a
Gaussian distributed perturbation, all odd $n$-point correlation
functions vanish and the higher even $n$-point correlators are
expressed in terms of sum of products of the two-point functions. In
this sense, the two-point correlation function of the scalar
perturbations has no information about the non-Gaussian distribution
of the primordial perturbation in the model. To explore the
non-Gaussianity of the density perturbations, we should study the
three-point correlation function.

In fact, every slow roll single field inflationary model with
canonical kinetic term which starts from Bunch-Davies vacuum state,
predicts a Gaussian distribution of the perturbations. Any deviation
from these conditions leads to large non-Gaussianity. In this regard
and to find the three-point correlation function, we should proceed
with the nonlinear perturbation theory and expand the action
\eqref{eq1} up to the cubic order in the small fluctuations. The
result, which is a complicated expression, is shown in the
Appendix~\ref{A}. Now, by using equation \eqref{eq13}, we eliminate
the perturbation parameter $\Psi$ in the third order action. By
presenting the new parameter ${\cal{X}}$ as
\begin{eqnarray}
\label{eq35}
\Phi=\frac{2(\kappa^{-2}+\dot{\phi}^{2}{\cal{N}}'-8H\dot{\alpha}){\Upsilon}}{2\kappa^{-2}H+6\dot{\phi}^{2}H{\cal{N}}'
-24H^{2}\dot{\alpha}}\hspace{1cm}
\nonumber\\
+\frac{a^{2}{\cal{X}}}{\kappa^{-2}+\dot{\phi}^{2}{\cal{N}}'-8H\dot{\alpha}}\,,\hspace{1cm}
\end{eqnarray}
and
\begin{equation}
\label{eq36}
\partial^{2}{\cal{X}}={\cal{W}}\dot{\Upsilon}\,,
\end{equation}
we find the cubic action, up to the leading order, as follows
\begin{eqnarray}
\label{eq37} S_{3}=\int dt\, d^{3}x\,\Bigg\{
\Bigg[\frac{3a^{3}}{\kappa^{2}c_{s}^{2}}\,
\Bigg(1-\frac{1}{c_{s}^{2}}\Bigg)
\Bigg(\epsilon-4\kappa^{2}H\dot{\alpha}\Bigg) \Bigg]{\Upsilon}\dot{\Upsilon}^{2}\nonumber\\
+\Bigg[\frac{a}{\kappa^{2}}\,\Bigg(\frac{1}{c_{s}^{2}}-1\Bigg)
\Bigg(\epsilon-4\kappa^{2}H\dot{\alpha}\Bigg)
\Bigg]{\Upsilon}\,(\partial{\Upsilon})^{2}\nonumber\\+\Bigg[\frac{a^{3}}{\kappa^{2}}\,\Bigg(\frac{1}{c_{s}^{2}\,H}\Bigg)
\Bigg(\frac{1}{c_{s}^{2}}-1\Bigg)\Bigg(\epsilon-4\kappa^{2}H\dot{\alpha}\Bigg)\Bigg]
\dot{\Upsilon}^{3}\nonumber\\-\Bigg[a^{3}\,\frac{2}{c_{s}^{2}}\Bigg(\epsilon-4\kappa^{2}H\dot{\alpha}\Bigg)\dot{\Upsilon}
(\partial_{i}{\Upsilon})(\partial_{i}{\cal{X}})\Bigg]\Bigg\}\,.\hspace{0.8cm}
\end{eqnarray}

At this point, by using the interacting picture, we can obtain the
three point correlation function. In the interaction picture, the
vacuum expectation value of the curvature perturbation $\Upsilon$
for the three-point operator is given by~\cite{Mal03,Che08,See05}
\begin{eqnarray}
\label{eq38} \langle
{\Upsilon}(\textbf{k}_{1})\,{\Upsilon}(\textbf{k}_{2})\,{\Upsilon}(\textbf{k}_{3})\rangle
=\hspace{4.9cm}\nonumber\\-i\int_{\tau_{i}}^{\tau_{f}}d \tau \, a\,
\langle0|
[{\Upsilon}(\tau_{f},\textbf{k}_{1})\,{\Upsilon}(\tau_{f},\textbf{k}_{2})\,{\Upsilon}(\tau_{f},\textbf{k}_{3}),
H_{int}(\tau)]|0\rangle,\nonumber\\
\end{eqnarray}
which is obtained in the conformal time interval between the
beginning and end of the inflation. In equation \eqref{eq38}, the
interacting Hamiltonian, $H_{int}$, is equal to the minus of the
lagrangian term of the cubic action. Given that the variation of the
coefficients in the brackets of the lagrangian \eqref{eq37} is
slower than the scale factor, to solve the integral of equation
\eqref{eq38}, we treat with these coefficients as constants.

Solving the integral \eqref{eq38}, gives the three-point correlation
function in the Fourier space as the following expression
\begin{eqnarray}
\label{eq39} \langle
{\Upsilon}(\textbf{k}_{1})\,{\Upsilon}(\textbf{k}_{2})\,{\Upsilon}(\textbf{k}_{3})\rangle
=\hspace{3.5cm}\nonumber\\
(2\pi)^{3}\delta^{3}(\textbf{k}_{1}+\textbf{k}_{2}+\textbf{k}_{3})B_{\Upsilon}(\textbf{k}_{1},\textbf{k}_{2},\textbf{k}_{3})\,,
\end{eqnarray}
where
\begin{equation}
\label{eq40}
B_{\Upsilon}(\textbf{k}_{1},\textbf{k}_{2},\textbf{k}_{3})=\frac{(2\pi)^{4}{\cal{A}}_{s}^{2}}{\prod_{i=1}^{3}
k_{i}^{3}}\,
{\cal{J}}_{\Upsilon}(\textbf{k}_{1},\textbf{k}_{2},\textbf{k}_{3}).
\end{equation}
The parameter ${\cal{J}}_{\Upsilon}$ is given by the following
expression
\begin{eqnarray}
\label{eq41}
{\cal{J}}_{\Upsilon}=\frac{3}{4}\Bigg(1-\frac{1}{c_{s}^{2}}\Bigg)
\zeta_{1}-\frac{1}{4}\Bigg(1-\frac{1}{c_{s}^{2}}\Bigg)\zeta_{2}\nonumber\\
+\frac{3}{2}\Bigg(\frac{1}{c_{s}^{2}}-1\Bigg)\zeta_{3},
\end{eqnarray}
where
\begin{equation}
\label{eq42}
\zeta_{1}=\frac{2}{K}\sum_{i>j}k_{i}^{2}\,k_{j}^{2}-\frac{1}{K^{2}}\sum_{i\neq
j}k_{i}^{2}\,k_{j}^{3}\,,
\end{equation}
\begin{equation}
\label{eq43}
\zeta_{2}=\frac{1}{2}\sum_{i}k_{i}^{3}+\frac{2}{K}\sum_{i>j}k_{i}^{2}\,k_{j}^{2}-\frac{1}{K^{2}}\sum_{i\neq
j}k_{i}^{2}\,k_{j}^{3}\,,
\end{equation}
\begin{equation}
\label{eq44}
\zeta_{3}=\frac{\left(k_{1}\,k_{2}\,k_{3}\right)^{2}}{K^{3}}\,,
\end{equation}
and
\begin{equation}
\label{eq45} K=k_{1}+k_{2}+k_{3}\,.
\end{equation}
Equation \eqref{eq40} shows the dependence of the three-point
correlator on the three momenta $k_{1}$ , $k_{2}$ and $k_{3}$.
Satisfying the translation invariance implies that these momenta
form a closed triangle which means that there should be a constraint
on the momenta as $k_{1}+k_{2}+k_{3}=0$. Discussion on the issue of
the shapes of the triangle is raised when one considers the
translation invariance~\cite{Bab04a,Kom05,Cre06,Lig06,Yad07}.
Depending on the amount of momenta, we are faced with different
shapes of triangle. For each shapes, there is a maximal signal in a
specific configuration of triangle. When there is a  maximal signal
in the squeezed limit with $k_{3}\ll k_{1}\simeq k_{2}$), the
corresponding shape is a local shape \cite{Gan94,Ver00,Wan00,Kom01}.
Another shape which has a signal at $k_{1}=k_{2}=k_{3}$ is called
equilateral triangle \cite{Bab04b}. A shape corresponding to folded
triangle is given by a linear combination of the equilateral and
orthogonal templates. The folded triangle is orthogonal
to the equilateral triangle \cite{Sen10} and has a maximal signal in
$k_{1}=2k_{2}=2k_{3}$ limit. There is a shape of non-Guassianity
with a positive signal at the equilateral configuration and a
negative signal at the folded configuration which is dubbed the
orthogonal configuration. To measure the amplitude of the
non-Gaussianity we use the dimensionless parameter $f_{_{NL}}$,
expressed as
\begin{equation}
\label{eq46}
f_{NL}=\frac{10}{3}\frac{{\cal{J}}_{\Upsilon}}{\sum_{i=1}^{3}k_{i}^{3}}\,.
\end{equation}
which is called ``nonlinearity parameter''.

In this paper, we investigate the equilateral and orthogonal
configurations of the non-Gaussianity. In this regard and to find
parameter ${\cal{J}}_{\Upsilon}$ in these configurations, we follow
Refs.~\cite{Fer09,Fel13,Byr14} and by considering
${\cal{J}}_{\Upsilon}=\sum_{i=1}^{3}{\cal{J}}_{\Upsilon}^{(i)}$, we
introduce quantity ${\cal{C}}$ by the following expression
\begin{equation}
\label{eq47}
{\cal{C}}(\check{B}_{\Upsilon}^{(i)}\check{B}_{\Upsilon}^{(j)})
=\frac{{\cal{Z}}(\check{B}_{\Upsilon}^{(i)}\check{B}_{\Upsilon}^{(j)})}{\sqrt{{\cal{Z}}(\check{B}_{\Upsilon}^{(i)}\check{B}_{\Upsilon}^{(i)})
{\cal{Z}}(\check{B}_{\Upsilon}^{(i)}\check{B}_{\Upsilon}^{(i)})}}
\end{equation}
where, $\check{B}_{\Upsilon}=\frac{B_{\Upsilon}}{{\cal{A}}_{s}^{2}}$
and
\begin{eqnarray}
\label{eq48}
{\cal{Z}}(\check{B}_{\Upsilon}^{(i)},\check{B}_{\Upsilon}^{(j)})\hspace{5.1cm}\nonumber\\=
\int
dk_{1}\,dk_{2}\,dk_{3}\,\check{B}_{\Upsilon}^{(i)}(k_{1},k_{2},k_{3})\check{B}_{\Upsilon}^{(j)}(k_{1},k_{2},k_{3})
w
\end{eqnarray}
in which $w=\frac{(k_{1}k_{2}k_{3})^{4}}{(k_{1}+k_{2}+k_{3})^{3}}$.
The momentum interval of integration is $0<k_{1}<\infty$,
$0\leq\frac{k_{2}}{k_{1}}<1$ and
$1-\frac{k_{2}}{k_{1}}\leq\frac{k_{3}}{k_{1}}\leq1$. Every two
shapes for which the constraint $|
{\cal{C}}(\check{B}_{\Upsilon}\check{B}'_{\Upsilon})|\simeq 0$ is
satisfied, are almost orthogonal. In this point, we introduce a
shape $\breve{\zeta}^{equil}$ as follows~\cite{Fel13,Noz14}
\begin{equation}
\label{eq49}
\breve{\zeta}^{equil}=-\frac{12}{13}\Big(3\zeta_{1}-\zeta_{2}\Big)\,.
\end{equation}
By using equations \eqref{eq40}, \eqref{eq41}, \eqref{eq47} and
\eqref{eq48}, it can be shown that the following shape is orthogonal
to \eqref{eq47}
\begin{equation}
\label{eq50}
\breve{\zeta}^{ortho}=\frac{12}{14-13\beta}\Big(\beta\big(3\zeta_{1}-\zeta_{2}\big)+3\zeta_{1}-\zeta_{2}\Big)\,,
\end{equation}
where $\beta\simeq 1.1967996$. Now, the bispectrum \eqref{eq41}, in
terms of the equilateral and orthogonal shapes basis is given by the
following expression
\begin{equation}
\label{eq51}
{\cal{J}}_{\Upsilon}={\cal{M}}_{1}\,\breve{\zeta}^{equil} +
{\cal{M}}_{2} \,\breve{\zeta}^{ortho}\,,
\end{equation}
where the parameters ${\cal{M}}_{1}$ and ${\cal{M}}_{2}$ are defined
as
\begin{equation}
\label{eq52}
{\cal{M}}_{1}=\frac{13}{12}\Bigg[\frac{1}{24}\bigg(1-\frac{1}{c_{s}^{2}}\bigg)\bigg(2+3\beta\bigg)
\Bigg]\,,
\end{equation}
and
\begin{equation}
\label{eq53}
{\cal{M}}_{2}=\frac{14-13\beta}{12}\Bigg[\frac{1}{8}\bigg(1-\frac{1}{c_{s}^{2}}\bigg)\Bigg]\,.
\end{equation}
The amplitudes of the non-Gaussianity in the equilateral and
orthogonal configurations, obtained from equations
\eqref{eq46}-\eqref{eq53}, are given by
\begin{equation}
\label{eq54}
f_{_{NL}}^{equil}=\frac{130}{36\sum_{i=1}^{3}k_{i}^{3}}\Bigg[\frac{1}{24}\bigg(1-\frac{1}{c_{s}^{2}}\bigg)\bigg(2+3\beta\bigg)
\Bigg]\breve{\zeta}^{equil}\,,
\end{equation}
and
\begin{equation}
\label{eq55}
f_{_{NL}}^{ortho}=\frac{140-130\beta}{36\,\sum_{i=1}^{3}k_{i}^{3}}\Bigg[\frac{1}{8}\bigg(1-\frac{1}{c_{s}^{2}}\bigg)
\Bigg]\breve{\zeta}^{ortho}\,.
\end{equation}
Now, we rewrite equations \eqref{eq54} and \eqref{eq55} in the limit
$k_{1}=k_{2}=k_{3}$ as
\begin{equation}
\label{eq56}
f_{_{NL}}^{equil}=\frac{325}{18}\Bigg[\frac{1}{24}\bigg(\frac{1}{c_{s}^{2}}-1\bigg)\bigg(2+3\beta\bigg)
\Bigg]\,,
\end{equation}
and
\begin{equation}
\label{eq57}
f_{_{NL}}^{ortho}=\frac{10}{9}\Big(\frac{65}{4}\beta+\frac{7}{6}\Big)\Bigg[\frac{1}{8}\bigg(1-\frac{1}{c_{s}^{2}}\bigg)
\Bigg]\,.
\end{equation}
This is because, there is a negative signal for the equilateral and
a positive signal for an orthogonal shape function at this limit.

Up to this point, the main equations describing the cosmological dynamics
of our setup, have been calculated. To investigate the cosmological
viability of the model, we compare the results of
our setup with the Planck2015 released data.

\begin{figure*}
\flushleft\leftskip0em{
\includegraphics[width=.32\textwidth,origin=c,angle=0]{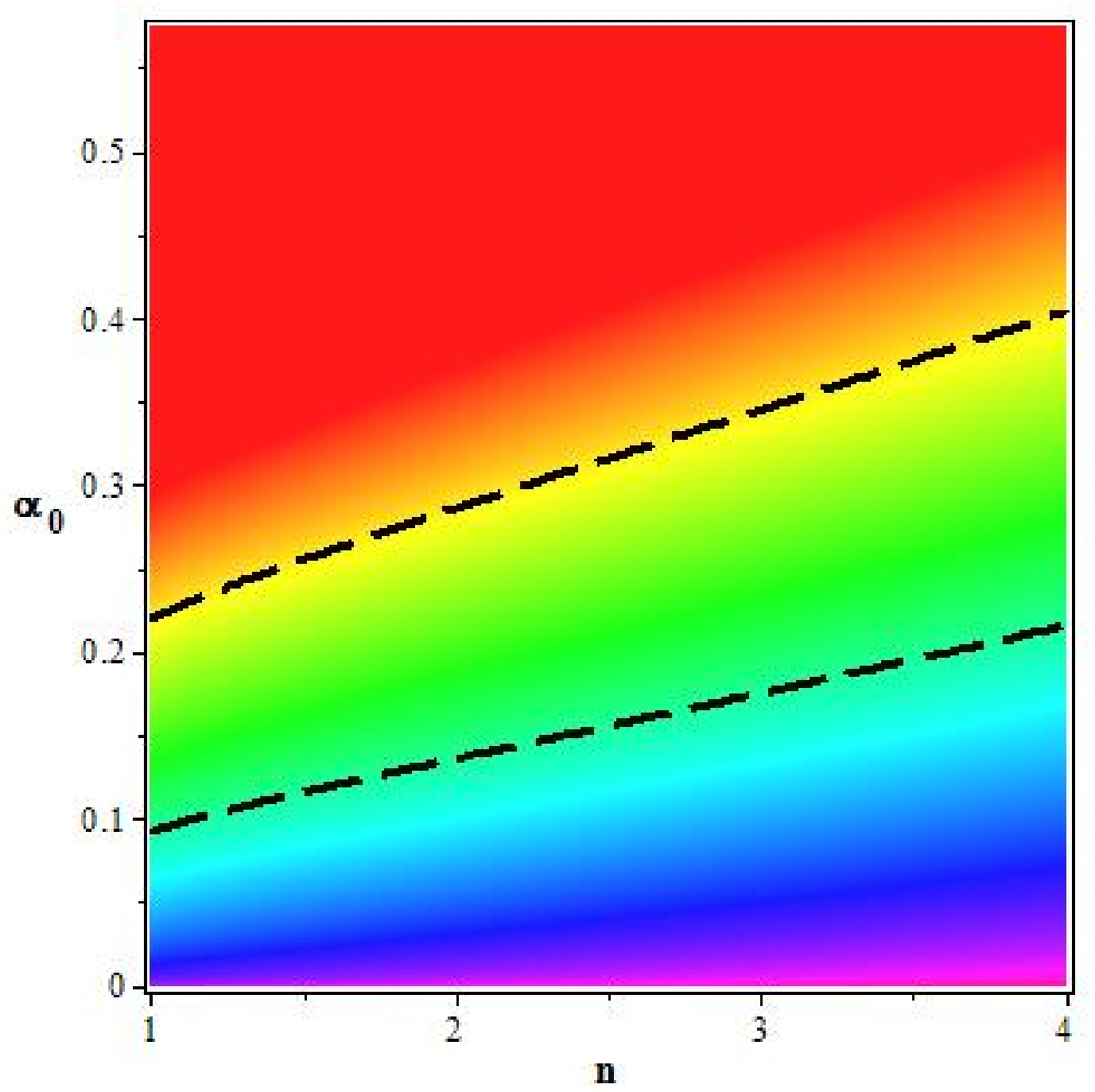}
\hspace{0.01cm}
\includegraphics[width=.1\textwidth,origin=c,angle=0]{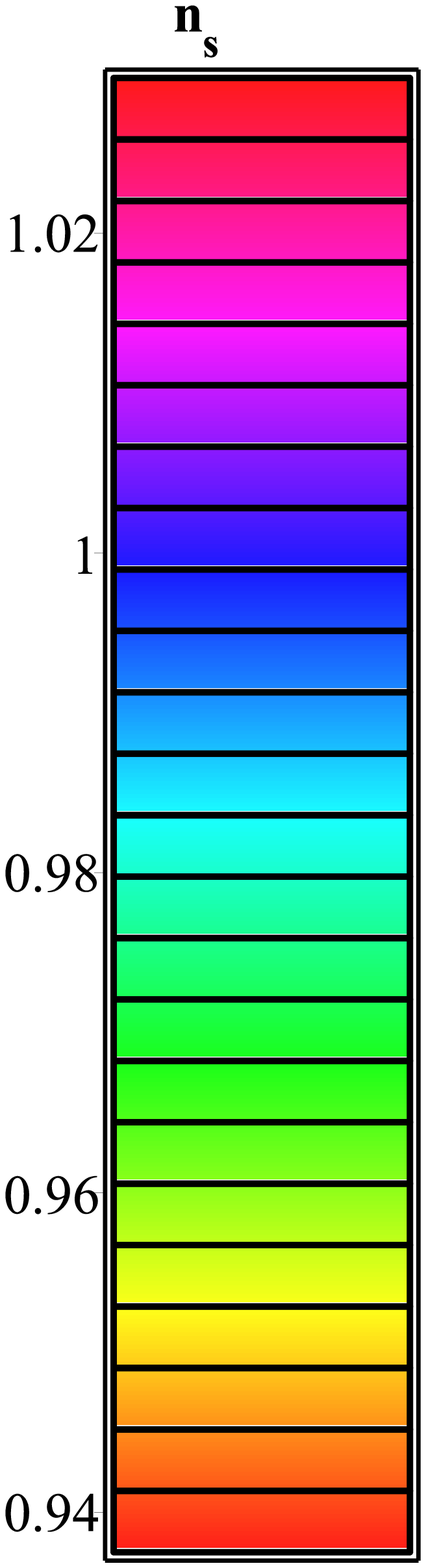}
\hspace{0.5cm}
\includegraphics[width=.33\textwidth,origin=c,angle=0]{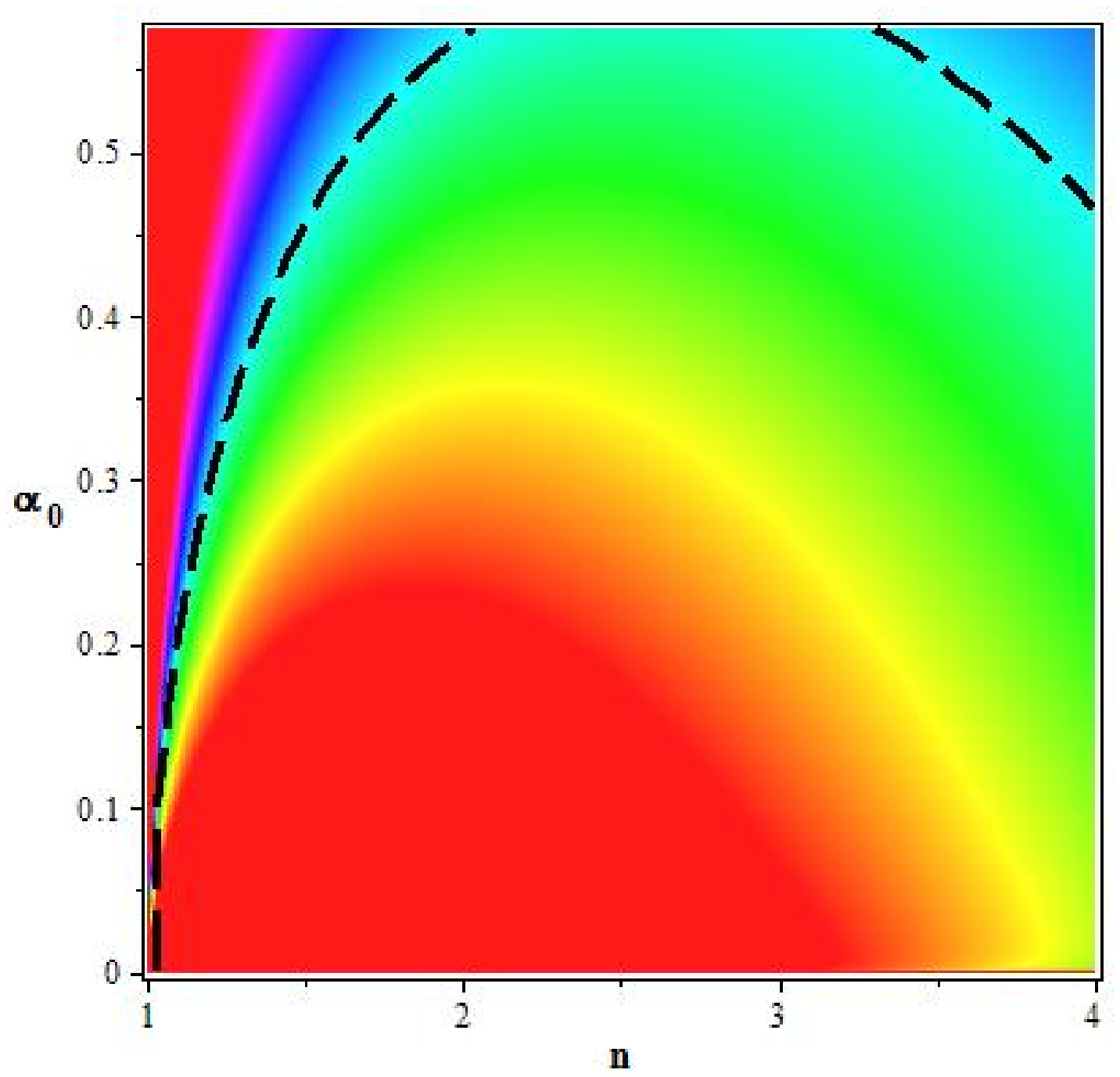}
\hspace{0.01cm}
\includegraphics[width=.1\textwidth,origin=c,angle=0]{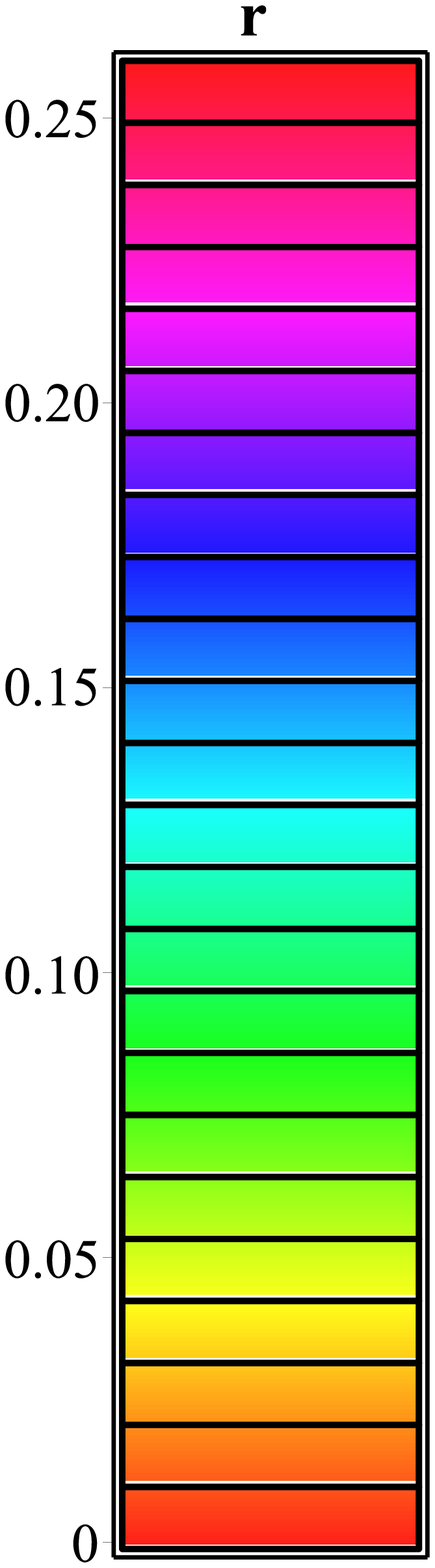}}
\caption{\label{fig1} The range of the parameters $\alpha_{0}$ and
$n$ which lead to the observationally viable values of the scalar
spectral index (left panel) and the tensor-to-scalar ratio (right
panel) for a NMDC inflationary
 model with GB coupling term. In the left panel, the region between the dashed curves
shows the Planck2015 constraints. In the right panel, the region
below the dashed curve shows the Planck2015 constraints. The panels
have been plotted with $N=60$.}
\end{figure*}

\begin{figure*}
\includegraphics[width=.38\textwidth,origin=c,angle=0]{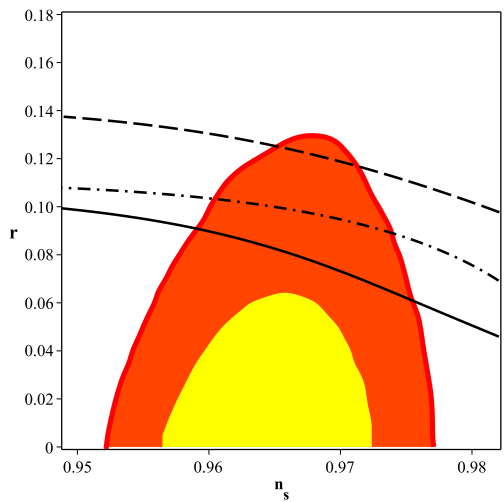}
 \caption{\label{fig2} Tensor-to-scalar ratio versus the scalar spectral index for a NMDC inflationary
 model with GB coupling term, in the background of Planck2015 TT, TE, EE+lowP data. The solid line is corresponding to $n=2$.
The dashed-dotted line is corresponding to $n=3$. The dashed line is
corresponding to $n=4$. The diagrams have been plotted with $N=60$.}
\end{figure*}

\begin{figure*}
\flushleft\leftskip0em{
\includegraphics[width=.35\textwidth,origin=c,angle=0]{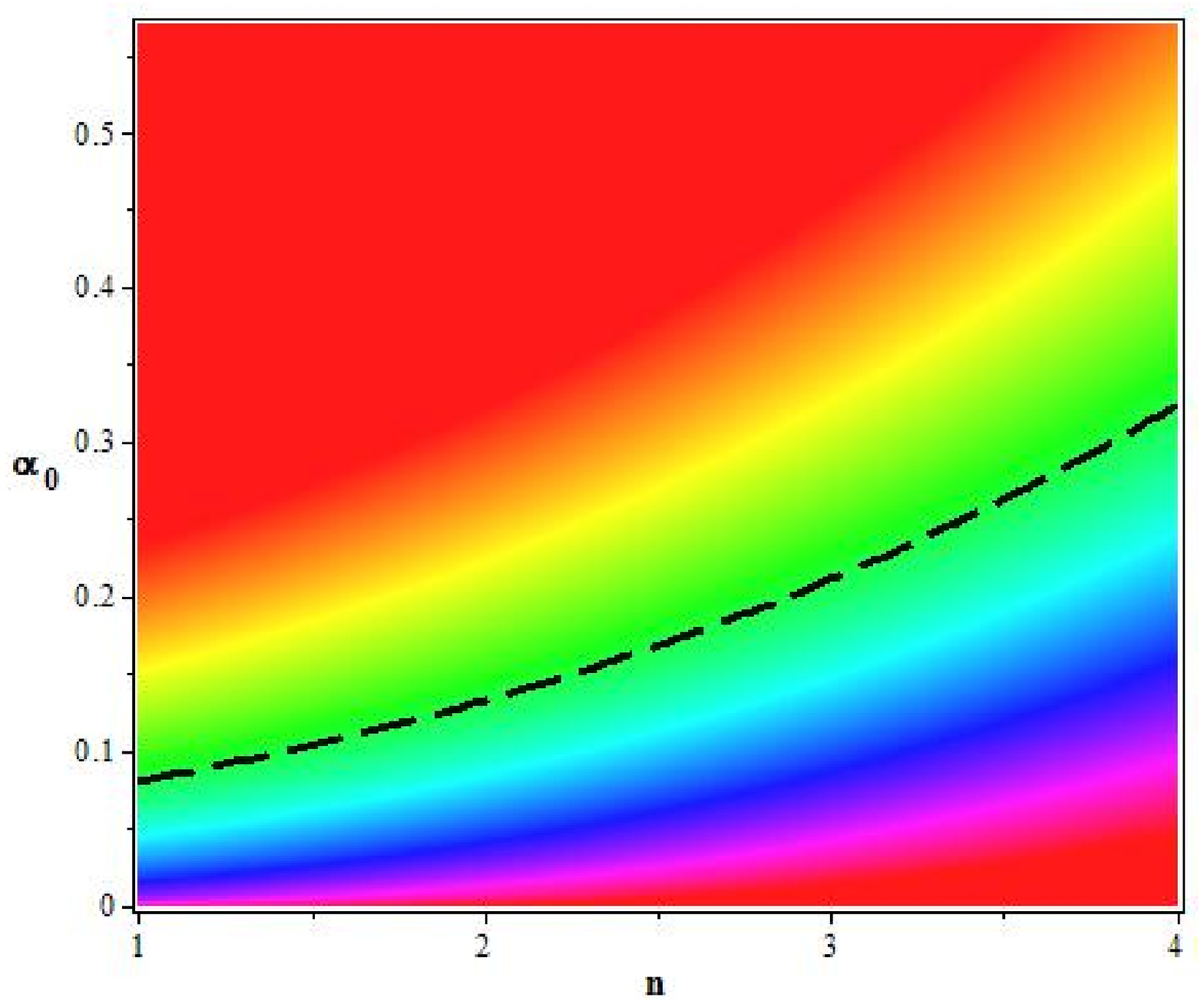}
\hspace{0.01cm}
\includegraphics[width=.1\textwidth,origin=c,angle=0]{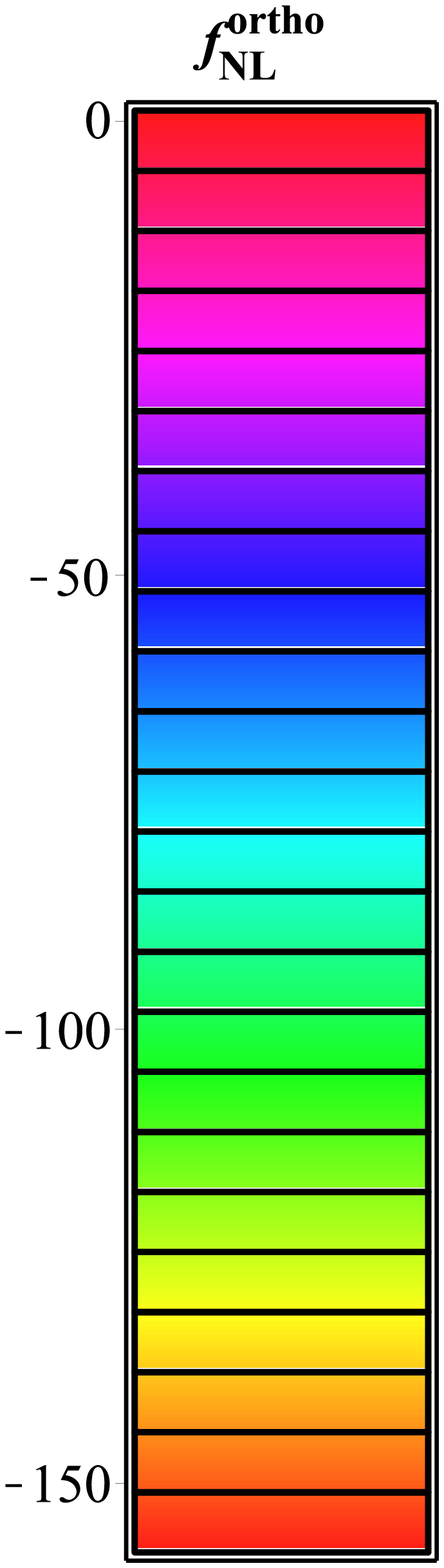}
\hspace{0.5cm}
\includegraphics[width=.35\textwidth,origin=c,angle=0]{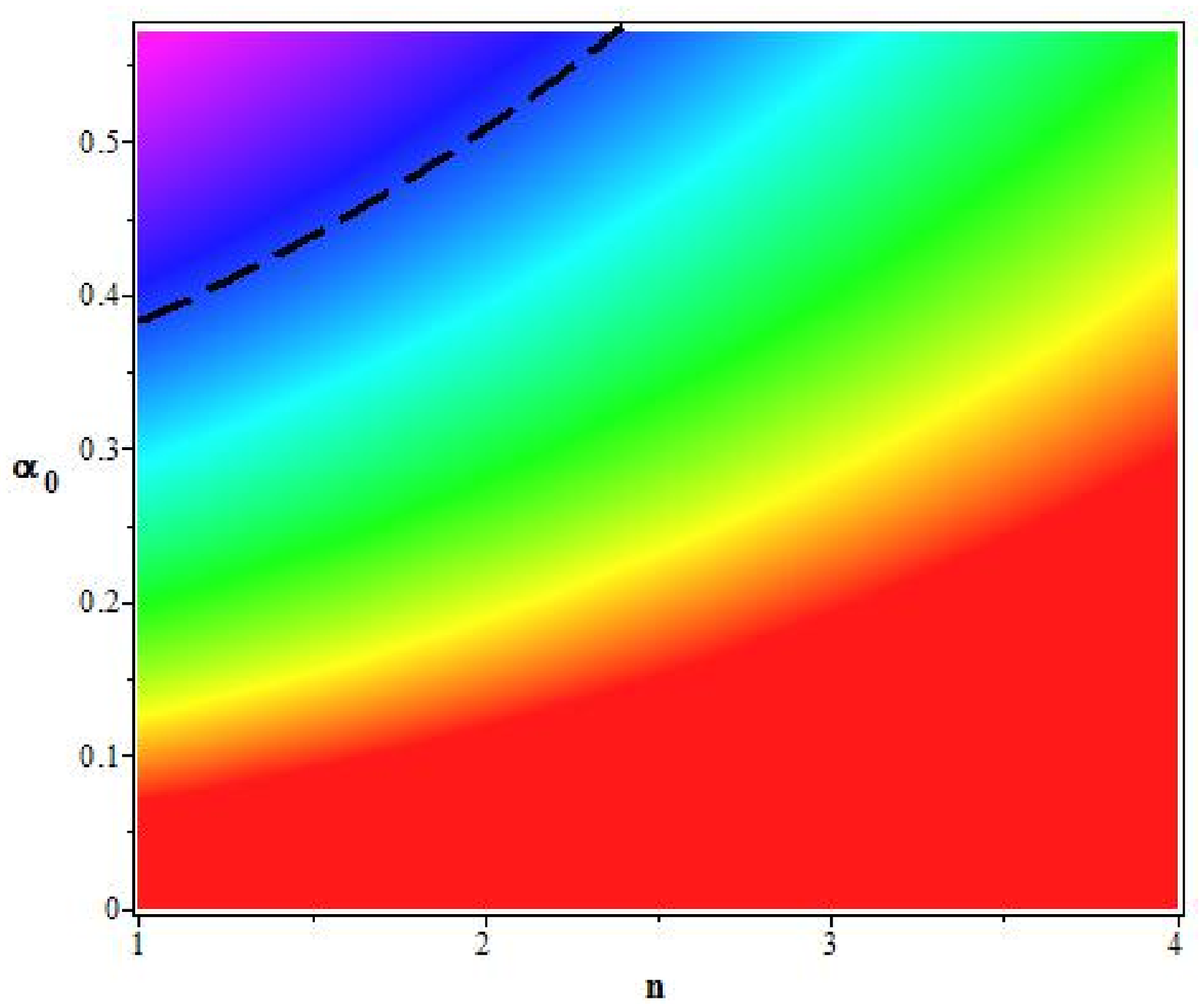}
\hspace{0.01cm}
\includegraphics[width=.095\textwidth,origin=c,angle=0]{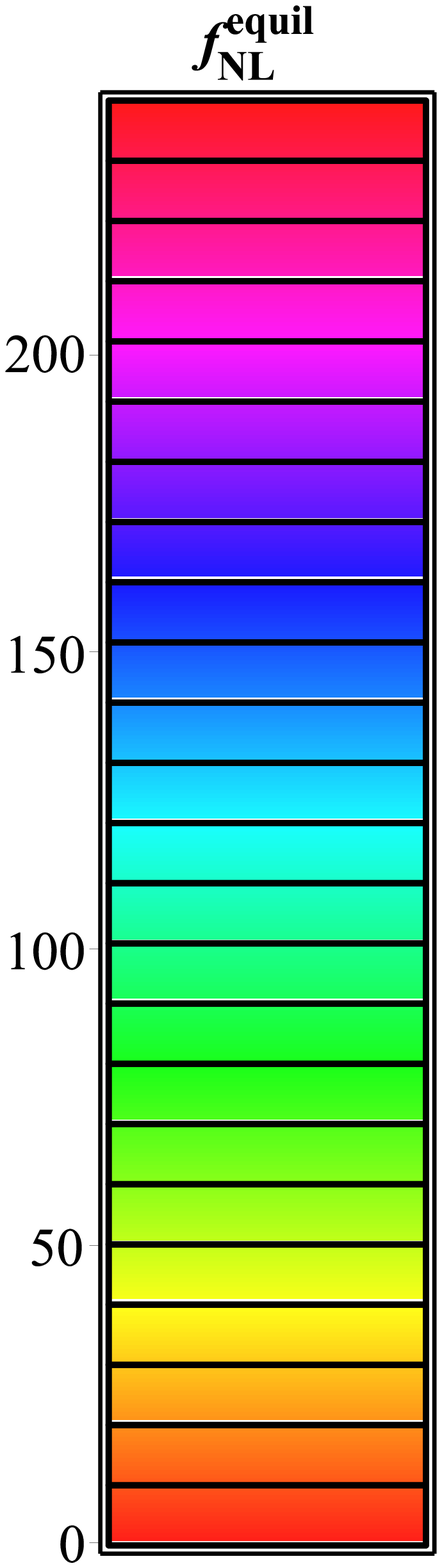}}
\caption{\label{fig3} The range of the parameters $\alpha_{0}$ and
$n$ which lead to the observationally viable values of the
orthogonal configuration of non-Gaussianity (left panel) and the
equilateral configuration of non-Gaussianity (right panel) for a
NMDC inflationary model with GB coupling term. In both panels, the
region below the dashed curve shows the Planck2015 constraints. The
panels have been plotted with $N=60$.}
\end{figure*}

\begin{figure*}
\includegraphics[width=.38\textwidth,origin=c,angle=0]{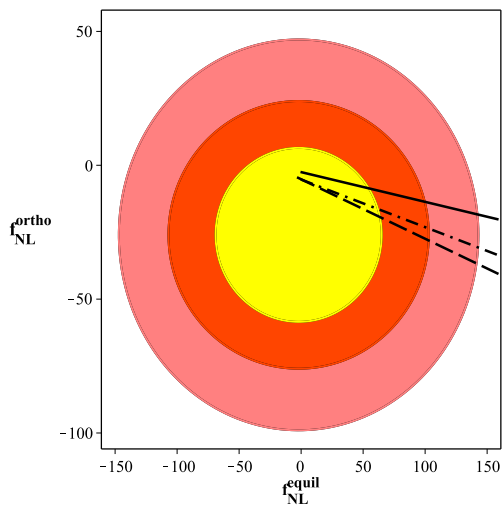}
 \caption{\label{fig4} Amplitude of the orthogonal configuration of
the non-Gaussianity versus the equilateral configuration for a NMDC
inflationary
 model with GB coupling term, in the background of Planck2015
TTT, EEE, TTE and EET data. The solid line is corresponding to
$n=2$. The dashed-dotted line is corresponding to $n=3$. The dashed
line is corresponding to $n=4$. The diagrams have been plotted with
$N=60$.}
\end{figure*}

\begin{table*}
\caption{\label{tab:1} The ranges of $\alpha_{_{0}}$ for which the
values of the inflationary parameters $r$, $n_{s}$, $f_{NL}^{ortho}$
and $f_{NL}^{equil}$ are compatible with the $95\%$ CL of the
Planck2015 dataset.}
\begin{tabular}{cccccc}
\\ \hline \hline&$n=2$&&
$n=3$&&$n=4$\\ \hline\\
$r-n_{s}$&  $0.138<\alpha_{0}<0.285$ &&$0.176<\alpha_{0}<0.34$& &$0.217<\alpha_{0}<0.41$\\\\
$f_{NL}^{ortho}-f_{NL}^{equil}$&  $\alpha_{0}<0.13$ &&$\alpha_{0}<0.21$&& $\alpha_{0}<0.32$\\\\
\hline\\\\
\end{tabular}
\end{table*}

\section{Observational Constraints}

In this section we turn our attention to the numerical analysis of
the model at hand and compare it with the observational data. In
this regard, we adopt the functions of scalar field, introduced in
action \eqref{eq1}, as ${\cal{N(\phi)}}\sim\frac{1}{2n}\phi^n$,
$\alpha(\phi)\sim\alpha_{0}\phi^n$ and $V(\phi)\sim
\frac{1}{n}\phi^n$. By these adoption of the general functions, we
do perform our numerical analysis and obtain some constraints on the
parameters space of the model. To this end, first of all, we should
substitute these functions into the integral of equation \eqref{eq8}
and solve this integral to find the value of the inflaton at the
horizon crossing of the physical scales. Next, we substitute the
obtained value of the field, $\phi_{hc}$, in equations \eqref{eq23},
\eqref{eq34}, \eqref{eq56} and \eqref{eq57}. In this way, we get the
scalar spectral index, tensor-to-scalar ratio, the amplitudes of the
orthogonal and equilateral configurations of the non-Gaussianity in
terms of $N$, $n$ and $\alpha_{0}$. Other constant parameters are
rescalled to the unity. Now , we are in the situation that we can
start exploring the cosmological observable parameters numerically.
We perform our computation by taking $N=60$ and observational
parameters are defined at $k_{0}=0.002 Mpc^{-1}$ (subscript $0$
means the value of $k$ at hc). The results are shown in Figures
~\ref{fig1}-\ref{fig4}. Figure~\ref{fig1} shows the ranges of the
parameters $\alpha_{0}$ and $n$ which lead to the observationally
viable values of the scalar spectral index (left panel) and the
tensor-to-scalar ratio (right panel). Figure shows that the NMDC
model in the presence of the GB effect, in some regions of the
$\alpha_{0}$ and $n$, is consistent with the Planck2015 dataset. By
increasing the value of $n$, the model with stronger coupling of GB
term is cosmologically viable. The interesting point is that this
model, for some values of $n$ and GB coupling term, predicts
blue-tilted spectrum ($n_{s}>1$). Also for this case by increasing
$n$, a larger region of $\alpha_{0}$ results the blue spectrum.
However, as the figure shows, a stronger GB coupling leads to
red-tilted spectrum. In figure~\ref{fig2}, we have plotted the
tensor-to-scalar ratio versus the scalar spectral index in the
background of the Planck2015 TT, TE, EE+lowP data. In this plot we
have set $n=2$, $3$ and $4$ and performed our analysis with these
values. Our analysis shows that with $n=2$, the model is consistent
with observational data if $0.138<\alpha_{0}<0.285$. With $n=3$, the
model has cosmological viability if $0.176<\alpha_{0}<0.34$. With
$n=4$, the model is compatible with Planck2015 dataset if
$0.217<\alpha_{0}<0.41$. The numerical analysis on the non-Gaussian
feature of the perturbation has been studied too. The results are
shown in figures~\ref{fig3} and~\ref{fig4}. Figure~\ref{fig3} shows
the range of the parameters $\alpha_{0}$ and $n$ in which the
orthogonal configuration of non-Gaussianity (left panel) and the
equilateral configuration of non-Gaussianity (right panel) are
observationally viable. Figure~\ref{fig4} shows the orthogonal
configuration versus the equilateral configuration of
non-Gaussianity in the background of Planck2015 TTT, EEE, TTE and
EET data. In this case also, we have considered three values of $n$
as $n=2$, $3$ and $4$. The corresponding diagrams are too close to
be distinguished. However, in the right panel of figure~\ref{fig4},
we have re-scaled and enlarged the plot in order to have more
resolution in diagrams. By studying the behavior of the
configurations of the non-Gaussianity, we have obtained some
constraint on the GB coupling term. With $n=2$, the observational
viability of the model is preserved if $\alpha_{0}<0.13$. With
$n=3$, the numerical analysis gives the constraint as
$\alpha_{0}<0.21$. With $n=4$, the model has consistency with
observational data if $\alpha_{0}<0.32$. The numerical results of
this section are summarized in table \ref{tab:1}.

\section{Summary}

In this paper, we have considered an inflationary model in which
both the inflaton scalar field and its derivatives are coupled
nonminimally to gravity. The scalar field is coupled to the
Gauss-Bonnet term and its derivatives are coupled to the Einstein
tensor term. These coupling have no problem with unitary. First of
all, we have found main equations of the inflationary dynamics in
this model. Then, we have focused on the primordial perturbation in
this model. In this regard, we have considered ADM formalism and
carried out our computations. We have expanded the action of the
model at hand, up to the second order in the perturbations and by
using the two point correlator. Then we have obtained the power
spectrum of the scalar perturbation and its spectral index. The
power spectrum of the tensor perturbation also has been computed. We
have considered the consistency relation in this setup and found
that in the presence of the Gauss-bonnet coupling, the consistency
relation gets modified. After that, by expanding the action up to
the third order in the perturbations and using the three point
correlation function we have studied the non-Gaussian feature of the
perturbations. In this regard, we have presented two shape functions
given by $\breve{\zeta}^{equil}$ and $\breve{\zeta}^{ortho}$ and by
that we have computed the amplitude of the non-Gaussianity in the
orthogonal and equilateral configurations in the limit
$k_{1}=k_{2}=k_{3}$, in which both configuration have peak. To test
the cosmological viability of this setup, in the section V, we have
compared the results of this model in confrontation with Planck2015
observational data. To this end, by adopting
${\cal{N(\phi)}}\sim\frac{1}{2n}\phi^n$,
$\alpha(\phi)\sim\alpha_{0}\phi^n$ and $V(\phi)\sim
\frac{1}{n}\phi^n$, we have plotted the region of $\alpha_{0}$ and
$n$ which lead to the observationally viable values of $n_{s}$ and
$r$. We have found that NMDC model with a GB coupling in some ranges
of the parameters space is consistent with Planck2015 observational
data. By increasing the value of $n$, more stronger coupling of the
Gauss-Bonnet term is needed to model be fitted with observation. A
notable point is that this model in some ranges of the parameters
predicts blue-tilted power spectrum. As $n$ increases, the model in
larger region of $n$ and $\alpha_{0}$ is consistent with the
Planck2015 dataset. As GB coupling becomes stronger, the model tends
to the red-tilted power spectrum. The tensor-to-scalar ratio versus
the scalar spectral index in the background of Planck2015 TT, TE,
EE+lowP data has been plotted and some constraints on the GB
coupling parameter $\alpha_{0}$ have been obtained. The non-Gaussian
feature of the perturbation also in this model has been studied. The
range of the cosmologically viable $n$ and $\alpha_{0}$ have been
plotted. By plotting the orthogonal configuration versus the
equilateral configuration of the non-Gaussianity in the background
of Planck2015 TTT, EEE, TTE and EET data, some constraints on the
model's parameters are obtained. This extended inflationary model in
some ranges of $n$ and $\alpha_{0}$ predicts large non-Gaussianity.
In fact, we expect large non-Gaussianity in this model as a result
of non-canonical kinetic term in the lagrangian. Also, in the
presence of the GB term, this model is not a single-field
inflationary model anymore. Consequently, this model has capability
to predict large non-Gaussianity.

\appendix
\section{The expanded action up to third order} \label{A}
\begin{widetext}
\begin{eqnarray}
S_{3}=\int dt\, d^{3}x\,
a^{3}\Bigg[\left(3\kappa^{-2}H^{2}-\frac{\dot{\phi}^{2}}{2}+30H^{2}\dot{\phi}^{2}{\cal{N}}'-80H^{3}\dot{\alpha}\right)\Psi^{3}
+\Psi^{2}\Bigg(\Big(\frac{3}{2}\dot{\phi}^{2}
-9\kappa^{-2}H^{2}+144H^{3}\dot{\alpha}-54H^{2}\dot{\phi}^{2}{\cal{N}}'\Big){\Upsilon}
\nonumber\\+\Big(-6\kappa^{-2}H
-36H\dot{\phi}^{2}{\cal{N}}'+144H^{2}\dot{\alpha}\Big)\dot{\Upsilon}
+\left(2\dot{\phi}^{2}{\cal{N}}'-16H\dot{\alpha}\right)\frac{\partial^{2}{\Upsilon}}{a^{2}}+\Big(-2\kappa^{-2}H
+12H\dot{\phi}^{2}{\cal{N}}'+48H^{2}\dot{\alpha}\Big)
\frac{\partial^{2}\Phi}{a^{2}}\Bigg)\nonumber\\+\Psi\Bigg(-a^{-2}\Big(2\kappa^{-2}H
+6\dot{\phi}^{2}{\cal{N}}'-24H^{2}\dot{\alpha}\Big)\partial_{i}{\Upsilon}\partial_{i}\Phi+9\Big(2\kappa^{-2}H
+6\dot{\phi}^{2}{\cal{N}}'-24H^{2}\dot{\alpha}\Big)\dot{\Upsilon}
{\Upsilon}+\frac{16\dot{\alpha}}{a^{2}}\dot{\Upsilon}\partial^{2}\Upsilon\nonumber\\-\frac{\kappa^{-2}+3\dot{\phi}^{2}
{\cal{N}}'-12H\dot{\alpha}}{2a^{4}}
\Big(\partial_{i}\partial_{j}\Phi
\partial_{i}\partial_{j}\Phi-\partial^{2}\Phi
\partial^{2}\Phi\Big)+\frac{16\dot{\alpha}}{a^{4}}\Big(\partial_{i}\partial_{j}\Phi\,
\partial_{i}\partial_{j}\Upsilon-\partial^{2}\Phi\,
\partial^{2}\Upsilon\Big)\nonumber-2\frac{\kappa^{-2}+\dot{\phi}^{2}{\cal{N}}'-8H\dot{\alpha}}{a^{2}}{\Upsilon}\partial^{2}{\Upsilon}\\-
\frac{2\kappa^{-2}H
+6\dot{\phi}^{2}{\cal{N}}'-24H^{2}\dot{\alpha}}{a^{2}}{\Upsilon}\partial^{2}\Phi-\frac{2\kappa^{-2}+6\dot{\phi}^{2}{\cal{N}}'-48H\dot{\alpha}}
{a^{2}}\dot{\Upsilon}\partial^{2}\Phi
-\frac{\kappa^{-2}+\dot{\phi}^{2}{\cal{N}}'-8H\dot{\alpha}}{a^{2}}(\partial{\Upsilon})^{2}
\nonumber\\+\Big(3\kappa^{-2}+9\dot{\phi}^{2}{\cal{N}}'+72H\dot{\alpha}
\Big)\dot{\Upsilon}^{2}\Bigg)+8\dot{\alpha}\dot{\Upsilon}^{3}+\frac{\kappa^{-2}-\dot{\phi}^{2}{\cal{N}}'-8\ddot{\alpha}}
{a^{2}}{\Upsilon}(\partial{\Upsilon})^{2}
-9\Big(\kappa^{-2}+\dot{\phi}^{2}{\cal{N}}'-8H\dot{\alpha}\Big)\dot{\Upsilon}^{2}\,{\Upsilon}\nonumber\\
+2\frac{\kappa^{-2}+\dot{\phi}^{2}{\cal{N}}'-8H\dot{\alpha}}{a^{2}}\dot{\Upsilon}\partial_{i}
{\Upsilon}\partial_{i}\Phi-\frac{8\dot{\alpha}}{a^{2}}\dot{\Upsilon}^{2}\partial^{2}\Phi+2\frac{\kappa^{-2}
+\dot{\phi}^{2}{\cal{N}}'-8H\dot{\alpha}}{a^{2}}\dot{\Upsilon}
{\Upsilon}\partial^{2}\Phi
\nonumber\\
+\frac{\frac{3}{2}\Big(\kappa^{-2}+\dot{\phi}^{2}{\cal{N}}'-8H\dot{\alpha}\Big){\Upsilon}-4\dot{\alpha}\dot{\Upsilon}}
{a^{4}}\Big(\partial_{i}\partial_{j}\Phi
\partial_{i}\partial_{j}\Phi-\partial^{2}\Phi
\partial^{2}\Phi\Big)
-2\frac{\kappa^{-2}+\dot{\phi}^{2}{\cal{N}}'-8H\dot{\alpha}}{a^{4}}\partial_{i}{\Upsilon}\partial_{i}\Phi\,
\partial^{2}\Phi\Bigg]\nonumber
\end{eqnarray}
\end{widetext}


\begin{thebibliography}{}

\bibitem {Gut81} A. Guth, \prd, \textbf{23}, 347 (1981).

\bibitem {Lin82} A. D. Linde, Phys. Lett. , \textbf{108 B}, 389
(1982)

\bibitem{Alb82} A. Albrecht and P. Steinhard, Phys. Rev. D, \textbf{48}, 1220
(1982).

\bibitem {Lin90} A. D. Linde, \emph{Particle Physics and Inflationary Cosmology}
(Harwood Academic Publishers, Chur, Switzerland, 1990).
[arXiv:hep-th/0503203].

\bibitem{Lid00a} A. Liddle and D. Lyth, \emph{Cosmological Inflation and Large-Scale
Structure}, (Cambridge University Press, 2000).

\bibitem {Lid97} J. E. Lidsey et al, Abney, Rev. Mod. Phys.,
\textbf{69}, 373, (1997).

\bibitem {Rio02} A. Riotto, [arXiv:hep-ph/0210162].

\bibitem{Lyt09} D. H. Lyth and A. R. Liddle, \emph{The Primordial Density
Perturbation} (Cambridge University Press, 2009).

\bibitem{Mal03} J. M. Maldacena, JHEP, \textbf{0305},
013, (2003).

\bibitem{Ade15a} P. A. R. Ade et al., [arXiv:1502.02114]
[astro-ph.CO].

\bibitem{Ade15b} P. A. R. Ade et al., [arXiv:1502.01589]
[astro-ph.CO].

\bibitem{Ade15c} P. A. R. Ade et al., [arXiv:1502.01592]
[astro-ph.CO].

\bibitem{Bar04} N. Bartolo, E. Komatsu, S. Matarrese and A. Riotto, Phys.Rept.,
 \textbf{402}, 103, (2004).

\bibitem{Che10} X. Chen, Adv. Astron. \textbf{2010}, 638979, (2010).

\bibitem{Fel11a} A. De Felice and S. Tsujikawa, Phys. Rev. D, \textbf{84},
083504, (2011).

\bibitem{Fel11b} A. De Felice and S. Tsujikawa, JCAP, \textbf{1104},
029, (2011).

\bibitem {Noz12} K. Nozari and N. Rashidi, Phys. Rev. D, \textbf{86}, 043505
(2012).

\bibitem {Noz13a} K. Nozari and N. Rashidi, Phys. Rev. D, \textbf{88}, 023519
(2013).

\bibitem{Noz13b} K. Nozari and N. Rashidi, Phys. Rev. D, \textbf{88},
084040 (2013).

\bibitem{Noz13c} K. Nozari and N. Rashidi, Astrophys. Space Sci. \textbf{350}, 339 (2014).

\bibitem {Bab04a} D. Babich, P. Creminelli and M. Zaldarriaga, JCAP, \textbf{0408},
009, (2004).

\bibitem{Che08} C. Cheung, P. Creminelli, A. L. Fitzpatrick, J. Kaplan and L.
Senatore, JHEP, \textbf{0803}, 014, (2008).

\bibitem{Ame93} L. Amendola, Phys. Lett. B, \textbf{301}, 175, (1993).

\bibitem{Ger10} C. Germani and A. Kehagias, Physical Review Letters, \textbf{105}, 011302, (2010).

\bibitem{Tsu12} S. Tsujikawa, Physical Review D, \textbf{85}, 083518, (2012).

\bibitem{Sad2013} H. M. Sadjadi and P. Goodarzi, JCAP, \textbf{02}, article 038, (2013).

\bibitem{Sar10} E. N. Saridakis and S. V. Sushkov, Physical Review
D, \textbf{81}, 083510, (2010).

\bibitem{Noz15} K. Nozari and N. Rashidi, Arxiv:1509.06240.

\bibitem{Noz14} K. Nozari, M.shoukrani and N. rashidi, Advances in High Energy Physics, \textbf{2014}, 343819,
http://dx.doi.org/10.1155/2014/343819, (2014)

\bibitem{Zwi85} B. Zwiebach, Phys. Lett. B, \textbf{156}, 315, (1985).

\bibitem{Bou85} D. G. Boulware and S. Deser, \prl, \textbf{55}, 2656, (1985).

\bibitem{Noj05} S. Nojiri, S. D. Odintsov and M. Sasaki, \prd, \textbf{71},
123509, (2005).

\bibitem{Noj07} S. Nojiri, S. D. Odintsov and P. V. Tretyakov, Phys. Lett. B, \textbf{651}, 224,
(2007).

\bibitem{Guo09} Z. K. Guo and D. J. Schwarz, \prd, \textbf{80}, 063523,
(2009).

\bibitem{Guo10} Z. K. Guo and D. J. Schwarz, \prd, \textbf{81}, 123520,
(2010).

\bibitem{Bro07} R. A. Brown, \emph{Brane world cosmology with Gauss-Bonnet and induced gravity terms},
(PhD Thesis, 2007), [arXiv:gr-qc/0701083].

\bibitem{Bam07} K. Bamba, Z. K. Guo and N. Ohta, Prog. Theor. Phys., \textbf{118},
879, (2007).

\bibitem{And07} K. Andrew, B. Bolen and C. A. Middleton, Gen. Rel. Grav., \textbf{39},
2061, (2007).

\bibitem{Noz08} K. Nozari and B. Fazlpour, JCAP, \textbf{0806}, 032,
(2008).

\bibitem {Noz09a} K. Nozari, and N. Rashidi, Int. J. Thoer. Phys., \textbf{48},
2800, (2009).

\bibitem {Noz09b} K. Nozari, and N. Rashidi, JCAP, \textbf{0909},
014, (2009).

\bibitem {Noz09c} K. Nozari, and N. Rashidi, Int. J. Mod. Phys. D, \textbf{19},
219, (2009).

\bibitem{Hin13} G. Hinshaw et al., Astrophys. J. Suppl. Ser. 208, 19 (2013).

\bibitem{Ade13} P. A. R. Ade et al., arXiv:1303.5082.

\bibitem{Car15} C. van de Bruck and C. Longden, arXiv:1512.04768.

\bibitem{Arn60} R. L. Arnowitt, S. Deser and C. W. Misner, Phys. Rev., \textbf{117},
1595, (1960).

\bibitem{Muk92} V. F. Mukhanov, H. A. Feldman, R. H. Brandenberger, Physics
Reports, \textbf{215}, 203, (1992).

\bibitem{Bau09} Daniel Baumann, [arXiv:0907.5424][hep-th].

\bibitem{Bar80} J. Bardeen, PRD, \textbf{22}, 1882, (1980).

\bibitem{See05} D. Seery and J. E. Lidsey, JCAP, \textbf{0506}, 003,
(2005).

\bibitem{Kom05} E. Komatsu, D. N. Spergel and B. D. Wandelt, Astrophys. J. \textbf{634},
14, (2005).

\bibitem{Cre06} P. Creminelli, A. Nicolis, L. Senatore, M. Tegmark and M.
Zaldarriaga, JCAP, \textbf{0605}, 004, (2006).

\bibitem{Lig06} M. Liguori, F. K. Hansen, E. Komatsu, S. Matarrese and A.
Riotto, Phys. Rev. D, \textbf{73}, 043505, (2006).

\bibitem{Yad07} A. P. S. Yadav, E. Komatsu and B. D. Wandelt, Astrophys. J., \textbf{664},
680,(2007).

\bibitem{Gan94} A. Gangui, F. Lucchin, S. Matarrese and S. Mollerach, ApJ, \textbf{430},
447,(1994).

\bibitem{Ver00} L. Verde, L. Wang, A. F. Heavens and M. Kamionkowski, MNRAS,
\textbf{313}, 141, (2000).

\bibitem{Wan00} L. Wang and M. Kamionkowski, Phys. Rev. D, \textbf{61},
063504, (2000).

\bibitem{Kom01} E. Komatsu and D. N. Spergel, Phys. Rev. D, \textbf{63},
063002, (2001).

\bibitem{Bab04b} D. Babich, P. Creminelli and M. Zaldarriaga, J. Cosmology
Astropart. Phys., \textbf{8}, 9, (2004).

\bibitem{Sen10} L. Senatore, K. M. Smith and M. Zaldarriaga, J. Cosmology
Astropart. Phys., \textbf{1}, 28,  (2010).

\bibitem{Fer09}
J. R. Fergusson and E. P. S. Shellard, \prd \textbf{80}, 043510,
(2009).

\bibitem{Fel13} A. De Felice, S. Tsujikawa, JCAP, \textbf{03}, 030,
(2013).

\bibitem{Byr14} C. T. Byrnes, [arXiv:1411.7002] [astro-ph.CO].


\end{thebibliography}
\end{document}